\documentclass[11pt]{article}

\usepackage[a4paper,margin=1in]{geometry}
\usepackage{amsmath,amssymb,amsfonts}
\usepackage{amsthm}
\usepackage{bm}
\usepackage{graphicx}
\usepackage{authblk}
\usepackage{microtype}
\usepackage{enumitem}
\usepackage{hyperref}

\setlist[itemize]{itemsep=2pt,topsep=4pt}
\setlist[enumerate]{itemsep=2pt,topsep=4pt}

\newcommand{\kb}{k_{\mathrm B}}

\theoremstyle{plain}
\newtheorem{theorem}{Theorem}
\newtheorem{proposition}{Proposition}
\newtheorem{corollary}{Corollary}

\theoremstyle{remark}
\newtheorem{remark}{Remark}

\title{Fluctuation Theorems from a Continuous-Time Markov Model of Information-Thermodynamic Capacity in Biochemical Signal Cascades}

\author[1,2,3]{\bfseries Tatsuaki Tsuruyama, Ph.D., M.D.}
\affil[1]{Department of Drug Discovery Medicine, Graduate School of Medicine, Kyoto, Japan}
\affil[2]{Department of Physics, Graduate School of Science, Tohoku University, Sendai, Japan}

\date{}

\begin{document}
\maketitle

\noindent ORCID: 0000-0002-3118-2826 \\
\noindent Email: \texttt{tsuruyam@kuhp.kyoto-u.ac.jp}

\begin{abstract}
Biochemical signaling cascades transmit intracellular information while dissipating energy under nonequilibrium conditions.
We model a cascade as a code string and apply information-entropy ideas to quantify an optimal transmission rate.
A time-normalized entropy functional is maximized to define a capacity-like quantity governed by a conserved multiplier.
To place the theory on a rigorous stochastic-thermodynamic footing, we formulate stepwise signaling as a continuous-time Markov jump
process with forward and reverse competing rates. The embedded jump chain yields well-defined transition probabilities that
justify time-scale-based expressions. Under local detailed balance, the log ratio of forward and reverse rates can be interpreted as
entropy production per event, enabling a trajectory-level derivation of detailed and integral fluctuation theorems.
We further connect the information-theoretic capacity to the mean dissipation rate and outline finite-time fluctuation structure via the
scaled cumulant generating function (SCGF) and Gallavotti--Cohen symmetry, including a worked example using MAPK/ERK timescales.
\end{abstract}

\section{Introduction}

Growth-factor signaling pathways are often interpreted as \emph{dynamical codes}: information is conveyed not only by the amplitude of pathway activity
but also by temporal features such as duration and decay \cite{PurvisLahav2013Cell}.
A paradigmatic and experimentally well-characterized example is the EGFR--MAPK/ERK cascade.
In PC12 cells, EGF and NGF can drive distinct ERK activation profiles (often transient versus sustained), and sustained activation has been linked to
differentiation-related outcomes, underscoring the functional importance of timescale \cite{Marshall1995Cell,Traverse1992BiochemJ,Cowley1994Cell}.
Because ERK phosphorylation time courses are routinely measured, one can extract effective activation/inactivation times (e.g.\ rise-to-peak, decay time, half-life),
providing a direct route from data to kinetic parameters.

This experimental accessibility motivates a quantitative question:
\emph{given measured activation and deactivation timescales, what are the limits and costs of information transmission through a biochemical cascade?}
Answering this requires a framework that simultaneously (i) treats a cascade as an information-processing channel with temporal constraints and
(ii) quantifies nonequilibrium irreversibility in a trajectory-resolved manner, because intracellular reactions fluctuate and dissipate energy.
While information-theoretic descriptions of signaling have a long tradition \cite{Shannon1948,WaltermannKlipp2011},
nonequilibrium stochastic thermodynamics provides complementary exact relations---fluctuation theorems---that constrain entropy production along single trajectories
\cite{Seifert2005,Crooks1999,Gaspard2004} and have growing biophysical relevance \cite{Collin2005,Sagawa2014}.

Here we develop an information--thermodynamic framework for stepwise biochemical signaling.
We first model a cascade as a code string in which reaction durations act as code lengths, and define a capacity-like quantity by maximizing entropy rate under a time budget.
We then place the construction on a rigorous continuous-time Markov jump process (CTMC) footing, derive detailed and integral fluctuation theorems for trajectory entropy production
via a forward/dual path-probability ratio, and outline the finite-time fluctuation structure through the SCGF and Gallavotti--Cohen symmetry.
Finally, we connect the abstract parameters to experimentally accessible MAPK/ERK time-course data (Sec.~\ref{subsec:mapk_param}) and provide a worked numerical illustration.

\section{Model and information-entropy formulation}

\subsection{A chain-reaction view of information transmission}

We model a signalling cascade as a sequence of $n$ molecular species $X_j$ ($1\le j\le n$), each of which can appear
in an inactive form $X_j$ or an active (modified) form $X_j^\ast$.
At the coarse-grained level, one step of the cascade consists of (i) interconversion within a layer and
(ii) activation of the next layer catalyzed by the active upstream species:
\[
X_j \rightleftarrows X_j^\ast,
\qquad
X_j^\ast + X_{j+1} \rightleftarrows X_j + X_{j+1}^\ast.
\]
The second reaction is written in catalytic form to emphasize that $X_j^\ast$ promotes the modification of $X_{j+1}$
without being consumed; alternative stoichiometries can be accommodated without changing the information-theoretic construction below.

Let the total pool of signalling molecules (copy number or total concentration) be
\begin{equation}
X=\sum_{j=1}^{n}(X_j+X_j^\ast).
\label{eq:X_total}
\end{equation}
We work with the corresponding fractions
\begin{equation}
p_j=\frac{X_j}{X},\qquad p_j^\ast=\frac{X_j^\ast}{X},
\label{eq:pj_def}
\end{equation}
which satisfy the normalization
\begin{equation}
\sum_{j=1}^{n}(p_j+p_j^\ast)=1.
\label{eq:normalization}
\end{equation}

\paragraph{Reaction durations as code lengths.}
To connect kinetics with coding, we associate to each layer $j$ a forward modification time $\tau_j>0$
and a reverse (demodification) time $\tau_j^\ast>0$.
Given the mixture $(p_j,p_j^\ast)$ of symbols in a long run, the total time budget (message duration) is
\begin{equation}
\tau = X\sum_{j=1}^{n}(p_j\tau_j+p_j^\ast\tau_j^\ast).
\label{eq:time_budget}
\end{equation}
\begin{remark}
Some presentations attach a sign to $\tau_j^\ast$ to mark direction.
Here we keep all durations positive (so they can be used consistently as code lengths)
and represent directionality through rate asymmetries and entropy production (Secs.~\ref{sec:ctmc}--\ref{sec:ft}).
\end{remark}

\subsection{Shannon entropy from symbol counts}

Consider a long signalling realisation in which the counts of each symbol are $\{X_j\}$ and $\{X_j^\ast\}$.
The number of distinct sequences consistent with these counts is the multinomial multiplicity
\begin{equation}
\Psi=\frac{X!}{\prod_{j=1}^{n}X_j!\prod_{j=1}^{n}X_j^\ast!}.
\label{eq:multinomial}
\end{equation}
Using Stirling's approximation, we defined the  entropy becomes
\begin{equation}
S=\log\Psi\simeq X\left(\sum_{j=1}^{n}p_j\log p_j+\sum_{j=1}^{n}p_j^\ast\log p_j^\ast\right),
\label{eq:Shannon}
\end{equation}
where $\log$ denotes the natural logarithm.
If one prefers bits rather than nats, use $S_{\rm bits}=S/\ln 2$.

\section{Entropy-rate maximization and capacity}
\label{sec:capacity}

To quantify an optimal signalling efficiency, we maximize $S$ subject to the normalization constraint
\eqref{eq:normalization} and the time-budget constraint \eqref{eq:time_budget}.
Introducing Lagrange multipliers $a$ and $b$, we consider
\begin{equation}
G = S + a\left[\sum_{j=1}^{n}(p_j+p_j^\ast)\right] + b\left[X\sum_{j=1}^{n}(p_j\tau_j+p_j^\ast\tau_j^\ast)\right].
\label{eq:G_functional}
\end{equation}
Differential \textit{G} with respect \textit{X} ,  $p_j$ , and $p_j^*$  with respect to  yields an exponential bias of symbol frequencies by their durations:
\begin{align}
p_j &= {\exp(-b\tau_j)}, \label{eq:pj_opt}\\
p_j^\ast &= {\exp(-b\tau_j^\ast)}, \label{eq:pjstar_opt}
\end{align}

enforces \eqref{eq:normalization}.  Substituting \eqref{eq:pj_opt}--\eqref{eq:pjstar_opt} into \eqref{eq:Shannon}, one finds that the maximal entropy scales
linearly with the available time budget (up to $\tau$-independent constants absorbed into $Z$),
\begin{equation}
S_{\max}\simeq b\,\tau.
\label{eq:Smax}
\end{equation}
We define the channel capacity as the maximum achievable entropy rate. Writing $K:=\log_2 e$ (nats-to-bits),
\begin{equation}
C:=\lim_{\tau\to\infty}K\frac{S_{\max}}{\tau}=K\,b.
\label{eq:capacity}
\end{equation}

\section{Continuous-time Markov formulation of stepwise signalling}
\label{sec:ctmc}

We now place the cascade on a rigorous stochastic-thermodynamic footing.

\subsection{CTMC dynamics and embedded transition probabilities}

We consider a nearest-neighbour CTMC on states $j\in\{1,\dots,n\}$.
Across each link $(j,j{+}1)$ we assign
\[
k(j\to j+1):=k_j^+>0,
\qquad
k(j+1\to j):=k_j^->0
\qquad (1\le j\le n-1).
\]
Define characteristic durations on each link
\begin{equation}
\tau_j := \frac{1}{k_j^+},\qquad \tau_j^\ast := \frac{1}{k_j^-}.
\label{eq:tau_from_rates}
\end{equation}

The escape rate from state $j$ is the sum of outgoing rates. With boundary conventions
$k_0^-:=0$ and $k_n^+:=0$, this reads
\[
\lambda_j := k(j\to j+1)+k(j\to j-1)=k_j^+ + k_{j-1}^-,
\qquad 1\le j\le n.
\]

Conditioned on the occurrence of a jump from state $j$, the embedded jump-chain probabilities are
\begin{align}
P(j+1\mid j)
&=\frac{k(j\to j+1)}{\lambda_j}
=\frac{k_j^+}{k_j^+ + k_{j-1}^-}
=\frac{1/\tau_j}{1/\tau_j + 1/\tau_{j-1}^\ast}
=\frac{\tau_{j-1}^\ast}{\tau_j+\tau_{j-1}^\ast},
\label{eq:P_forward}
\\
P(j-1\mid j)
&=\frac{k(j\to j-1)}{\lambda_j}
=\frac{k_{j-1}^-}{k_j^+ + k_{j-1}^-}
=\frac{1/\tau_{j-1}^\ast}{1/\tau_j + 1/\tau_{j-1}^\ast}
=\frac{\tau_j}{\tau_j+\tau_{j-1}^\ast}.
\label{eq:P_backward}
\end{align}
Thus time-scale ratios enter transition probabilities through competing rates while preserving nonnegativity and conditional normalization.
(At boundaries, interpret the formula via the rate form; e.g.\ $P(2\mid 1)=1$ since $k_0^-=0$.)

\subsection{Log-ratio observable on each link}

For each link $(j,j{+}1)$ define the log-ratio (edge affinity)
\begin{equation}
z_j := \log\frac{k_j^-}{k_j^+}.
\label{eq:zj_def}
\end{equation}
Using Eq.\ \eqref{eq:tau_from_rates},
\begin{equation}
z_j = \log\frac{\tau_j}{\tau_j^\ast}.
\label{eq:zj_tau}
\end{equation}
Forward bias $k_j^+>k_j^-$ implies $z_j<0$.

\section{Fluctuation theorems and capacity--irreversibility link}
\label{sec:ft}

\subsection{Dual (adjoint) dynamics}

Assume the forward CTMC admits a stationary distribution $\pi^{\mathrm{ss}}=\{\pi^{\mathrm{ss}}_j\}_{j=1}^n$.
Define the adjoint (dual) rates by
\begin{equation}
k^\dagger(i\to j):=\frac{\pi^{\mathrm{ss}}_j}{\pi^{\mathrm{ss}}_i}\,k(j\to i)
\qquad (i\neq j).
\label{eq:dual_rates}
\end{equation}
A key consequence is that the escape rates coincide; hence the exponential waiting-time densities coincide:
\begin{equation}
\lambda^\dagger_j=\lambda_j,
\qquad\text{and}\qquad
f^\dagger_j(\Delta t)=f_j(\Delta t)=\lambda_j e^{-\lambda_j\Delta t}.
\label{eq:waiting_time}
\end{equation}

\subsection{Path probability and trajectory entropy production}

A trajectory $\omega$ over duration $t$ is specified by jump times $0<t_1<\cdots<t_m<t$ and visited states
$j_0\to j_1\to\cdots\to j_m$.
Define embedded probabilities
\[
P(j_{i+1}\mid j_i)=\frac{k(j_i\to j_{i+1})}{\lambda_{j_i}},
\qquad
P^\dagger(j_{i+1}\mid j_i)=\frac{k^\dagger(j_i\to j_{i+1})}{\lambda^\dagger_{j_i}}.
\]
The forward path probability density is
\begin{equation}
\mathcal{P}_F(\omega)
=\pi_0(j_0)\prod_{i=0}^{m-1}\Big[P(j_{i+1}\mid j_i)\,f_{j_i}(t_{i+1}-t_i)\Big],
\label{eq:path_forward}
\end{equation}
and the dual path probability density for the \emph{same} trajectory is
\begin{equation}
\mathcal{P}_\dagger(\omega)
=\pi_0(j_0)\prod_{i=0}^{m-1}\Big[P^\dagger(j_{i+1}\mid j_i)\,f^\dagger_{j_i}(t_{i+1}-t_i)\Big].
\label{eq:path_dual}
\end{equation}
By Eq.~\eqref{eq:waiting_time}, waiting-time factors cancel in the log ratio below.

Define the trajectory entropy production (irreversibility) by
\begin{equation}
\Sigma_t(\omega):=\kb\log\frac{\mathcal{P}_F(\omega)}{\mathcal{P}_\dagger(\omega)}.
\label{eq:Sigma_def}
\end{equation}
Using Eq.~\eqref{eq:waiting_time}, one obtains
\[
\frac{\Sigma_t(\omega)}{\kb}
=\sum_{i=0}^{m-1}\log\frac{k(j_i\to j_{i+1})}{k^\dagger(j_i\to j_{i+1})}
=\sum_{i=0}^{m-1}\log\frac{k(j_i\to j_{i+1})\,\pi^{\mathrm{ss}}_{j_i}}{k(j_{i+1}\to j_i)\,\pi^{\mathrm{ss}}_{j_{i+1}}}.
\]

\subsection{Main results: DFT, IFT, second law}

\begin{theorem}[Detailed fluctuation theorem (DFT), forward/dual form]
Let $\Sigma_t$ be defined by Eq.~\eqref{eq:Sigma_def}.
Denote by $P_{F,t}(\Sigma)$ the probability density of $\Sigma_t$ under the forward dynamics,
and by $P^\dagger_t(\Sigma)$ the probability density of $-\Sigma_t$ under the dual dynamics.
Then
\begin{equation}
\frac{P_{F,t}(\Sigma)}{P^\dagger_t(-\Sigma)}=\exp\!\left(\frac{\Sigma}{\kb}\right).
\label{eq:DFT}
\end{equation}
\end{theorem}

\begin{corollary}[Integral fluctuation theorem (IFT)]
\begin{equation}
\left\langle \exp\!\left(-\frac{\Sigma_t}{\kb}\right)\right\rangle_{F}=1.
\label{eq:IFT}
\end{equation}
\end{corollary}

\begin{corollary}[Second-law inequality]
By Jensen's inequality,
\begin{equation}
\langle \Sigma_t\rangle_{F} \ge 0,
\qquad
\sigma:=\liminf_{t\to\infty}\frac{1}{t}\langle \Sigma_t\rangle_{F} \ge 0.
\label{eq:second_law}
\end{equation}
\end{corollary}

\section{Finite-time fluctuations: SCGF and Gallavotti--Cohen symmetry}

To characterize finite-time fluctuations and connect to large-deviation structure, consider the scaled cumulant generating
function (SCGF) of $\Sigma_t$:
\begin{equation}
\psi(\lambda):=\lim_{t\to\infty}\frac{1}{t}\log\left\langle \exp\!\left(-\lambda\frac{\Sigma_t}{\kb}\right)\right\rangle.
\label{eq:SCGF}
\end{equation}
For Markov jump processes, $\psi(\lambda)$ is given by the dominant eigenvalue of a \emph{tilted generator} (see, e.g.,
\cite{LebowitzSpohn1999,SeifertReview2012}).

Let the state space be the cascade steps. Denote by $k_{ij}$ the transition rate from $i$ to $j$ and define the entropy increment
$a_{ij}:=\log\frac{k_{ij}}{k_{ji}}$.
The tilted generator $(\mathcal L_\lambda)$ has off-diagonal entries
\[
(\mathcal L_\lambda)_{ij}=k_{ij}\exp\big(-\lambda a_{ij}\big)\quad (i\ne j),
\]
and diagonal entries chosen so each row sums to zero.

\begin{proposition}[Gallavotti--Cohen (GC) symmetry of the SCGF]
For steady-state entropy production in a Markov jump process,
\begin{equation}
\psi(\lambda)=\psi(1-\lambda).
\label{eq:GC}
\end{equation}
Equivalently, the large-deviation rate function $I(s)$ satisfies $I(s)-I(-s)=s$.
\end{proposition}

\subsection{Practical implication / Parameterization by MAPK time courses}
\label{subsec:mapk_param}

To connect the abstract stepwise time parameters with experimentally accessible kinetics,
we interpret $\tau_{\rm on}$ and $\tau_{\rm off}$ as effective activation and inactivation waiting times extracted from
time-course measurements of ERK phosphorylation dynamics.
With $k_+ \approx 1/\tau_{\rm on}$ and $k_- \approx 1/\tau_{\rm off}$, the dimensionless ratio
$\ln(\tau_{\rm off}/\tau_{\rm on})=\ln(k_+/k_-)$ provides an experimentally estimable proxy for the driving force that
controls irreversibility and dissipation in our framework (Secs.~\ref{sec:ctmc}--\ref{sec:ft}).

When half-times $t_{1/2}$ are reported, we use
$k_{\rm off}=\ln 2/t_{1/2}$ and $\tau_{\rm off}=1/k_{\rm off}=t_{1/2}/\ln 2$.

\begin{table}[t]
\centering
\caption{Example literature-based parameter values for MAPK/ERK timescales in an effective two-state step model.
When a half-time is available (explicitly or implicitly), $\tau_{\rm off}$ is converted to the exponential time constant:
$\tau_{\rm off}=t_{1/2}/\ln2$.}
\label{tab:mapk_params}
\begin{tabular}{p{4.2cm} p{2.1cm} p{2.4cm} p{2.2cm} p{2.3cm} p{2.0cm}}
\hline
System / condition &
$\tau_{\rm on}$ (min) &
$\tau_{\rm off}$ (min) &
$k_{\rm off}$ (min$^{-1}$) &
$\ln(\tau_{\rm off}/\tau_{\rm on})$ &
Source \\
\hline
EGF$\to$ERK transient response (peak $\sim$5 min; $\lesssim 50\%$ by 30 min) &
5 &
$30/\ln2 \approx 43.3$ &
$\ln2/30 \approx 0.0231$ &
$\ln(43.3/5)\approx 2.16$ &
\cite{Zhou2011PLOS} \\
\hline
HeLa; ERK inactivation after MEK inhibition (EGF-stimulated) &
-- &
$1.90/\ln2\approx 2.74$ &
$\ln 2/1.90 \approx 0.365$ &
-- &
\cite{Finch2012CellSig} \\
\cline{1-6}
HeLa; ERK inactivation after MEK inhibition (PDBu-stimulated) &
-- &
$0.92/\ln2\approx 1.33$ &
$\ln 2/0.92 \approx 0.753$ &
-- &
\cite{Finch2012CellSig} \\
\cline{1-6}
Model reduction used in \cite{Finch2012CellSig} (single dephosphorylation step) &
-- &
$\tau_{\rm off}=1/0.42 \approx 2.38$ &
$0.42$ &
-- &
\cite{Finch2012CellSig} \\
\hline
\end{tabular}
\end{table}

Assuming local detailed balance, one obtains an explicit dissipation estimate from timescales:
\begin{equation}
\Delta Q \approx \kb T \ln\frac{k_+}{k_-}
\approx \kb T \ln\frac{\tau_{\rm off}}{\tau_{\rm on}}.
\label{eq:dQ_from_times}
\end{equation}

\subsection{Worked example using MAPK/ERK timescales (Table~\ref{tab:mapk_params})}
\label{subsec:worked_example_mapk}

We replace the SCGF plot by a worked calculation based on MAPK/ERK time scales (Table~\ref{tab:mapk_params}).
Identify $k_{+}\approx 1/\tau_{\rm on}$ and $k_{-}\approx 1/\tau_{\rm off}$, and define the affinity
\begin{equation}
a:=\log\frac{k_{+}}{k_{-}}=\log\frac{\tau_{\rm off}}{\tau_{\rm on}}.
\label{eq:affinity_a}
\end{equation}

\paragraph{Example: EGF$\to$ERK transient response.}
Using $\tau_{\rm on}=5~{\rm min}$ and $\tau_{\rm off}=30/\ln2\simeq 43.3~{\rm min}$ from Table~\ref{tab:mapk_params}, we obtain
\[
k_{+}=0.20~{\rm min}^{-1},\qquad
k_{-}=0.0231~{\rm min}^{-1},\qquad
a=\log(43.3/5)\simeq 2.16.
\]
Thus the entropy production per effective step (in units of $k_{\mathrm B}$) is $\Sigma_{\rm step}/k_{\mathrm B}\approx a$.
Under local detailed balance, the corresponding dissipated heat per step is $\Delta Q \approx k_{\mathrm B}T\,a$.
At physiological temperature $T=310~{\rm K}$ this gives
\[
\Delta Q \approx 9.2\times 10^{-21}~{\rm J}\approx 9.2~{\rm pN\,nm}.
\]

If we further take an effective cycle time $\tau_{\rm cyc}\approx \tau_{\rm on}+\tau_{\rm off}=48.3~{\rm min}$,
an order-of-magnitude estimate of the mean entropy production rate is
\[
\frac{\sigma}{k_{\mathrm B}}\approx \frac{a}{\tau_{\rm cyc}}\simeq 4.5\times 10^{-2}~{\rm min}^{-1},
\qquad
\sigma_{\rm bits}\approx \frac{a/\ln 2}{\tau_{\rm cyc}}\simeq 6.4\times 10^{-2}~{\rm bits\,min}^{-1}.
\]

\paragraph{Fluctuation-theorem implication (finite-$N$ estimate).}
For a coarse-grained description with $N$ effective steps, the detailed fluctuation theorem \eqref{eq:DFT} implies
\[
\frac{P(-Nk_{\mathrm B}a)}{P(+Nk_{\mathrm B}a)}=\exp(-Na).
\]
With $a\simeq 2.16$, already $N=10$ yields $\exp(-Na)\approx 4.3\times 10^{-10}$, illustrating strong irreversibility at modest step counts.

\section{Discussion}

We modeled a signalling cascade as a code sequence in information science and derived an entropy-rate maximization principle.
The resulting capacity-like quantity $C$ is governed by a conserved multiplier $b$ and quantifies optimal information transmission efficiency.

A main development is the rigorous stochastic-thermodynamic formulation: by casting stepwise signalling as a CTMC,
transition probabilities expressed through time scales are justified via competing rates, and the fluctuation theorem follows from the path-probability ratio
(Eqs.~\eqref{eq:path_forward}--\eqref{eq:DFT}). Under local detailed balance, the log rate ratio can be interpreted as entropy production per event,
providing an experimentally interpretable bridge from measured activation/inactivation time scales to dissipation (Eq.~\eqref{eq:dQ_from_times}).

\paragraph{Units and the $C=\sigma$ identification.}
The capacity defined in Sec.~\ref{sec:capacity} is an \emph{entropy rate}.
In \emph{nats} (i.e.\ using the natural logarithm), the entropy production rate $\sigma$ is naturally measured in units of $\kb$ per unit time,
and the maximized entropy rate $S_{\max}/\tau$ is measured in nats per unit time.
Thus, in nats (equivalently, in $\kb$-scaled entropy units), one can identify the two rates up to the modeling/coarse-graining convention:
\[
\frac{1}{t}\left\langle \frac{\Sigma_t}{\kb}\right\rangle \ \leftrightarrow\ \,\frac{S_{\max}}{\tau} \ (= b),
\]
so that $\sigma$ corresponds to $-b$ in the same entropy units.
If one reports capacity in \emph{bits} per unit time, then $C=K(-b)$ with $K=\log_2 e$ (Eq.~\eqref{eq:capacity});
the corresponding entropy production rate in bits is $\sigma_{\rm bits}=(\sigma/\kb)/\ln 2$.

Finally, we outlined the finite-time fluctuation structure via SCGF and GC symmetry and provided a worked numerical example based on MAPK/ERK time-course data.
Future work should refine biochemical mappings (including chemical work from ATP hydrolysis and coarse-graining effects) and test predictions using time-resolved measurements.

\section{Conclusion}

A biochemical signalling cascade can be modeled as a code string to analyze information transmission.
Maximizing a time-normalized entropy functional yields a capacity-like measure.
By formulating signalling as a continuous-time Markov jump process and adopting local detailed balance, we derive detailed and integral fluctuation theorems for trajectory entropy production.
In entropy units, the information-theoretic capacity corresponds to a dissipation-rate scale, and the finite-time fluctuation structure is characterized by SCGF and GC symmetry.
This provides a mathematically rigorous and experimentally interpretable information-thermodynamic framework for biochemical signal transduction.

\section*{Acknowledgments}
This research was funded by a Grant-in-Aid from the Ministry of Education, Culture, Sports, Science, and Technology of Japan
(Synergy of Fluctuation and Structure: Quest for Universal Laws in Non-Equilibrium Systems, P2013-201 Grant-in-Aid for Scientific
Research on Innovative Areas, MEXT, Japan).

\section*{Conflicts of interest / Competing interests}
Not applicable.

\section*{Data availability}
Not applicable.

\section*{Ethical approval}
Not applicable.



\begin{thebibliography}{99}

\bibitem{Brillouin2013}
L.~Brillouin,
\emph{Science and Information Theory}, 2nd ed.,
Dover Publications (2013).

\bibitem{WaltermannKlipp2011}
C.~Waltermann and E.~Klipp,
\emph{Information theory based approaches to cellular signaling},
Biochim. Biophys. Acta \textbf{1810}(10), 924--932 (2011).
doi:10.1016/j.bbagen.2011.07.009

\bibitem{TsuruyamaEntropy2018}
T.~Tsuruyama,
\emph{Information Thermodynamics of the Cell Signal Transduction as a Szilard Engine},
Entropy \textbf{20}(4), 224 (2018).
doi:10.3390/e20040224

\bibitem{KisoFarneTsuruyama2022}
K.~Kiso-Farn\`e and T.~Tsuruyama,
\emph{Epidermal growth factor receptor cascade prioritizes the maximization of signal transduction},
Scientific Reports \textbf{12}, 16950 (2022).
doi:10.1038/s41598-022-20663-0

\bibitem{Shannon1948}
C.~E.~Shannon,
\emph{A mathematical theory of communication},
Bell Syst. Tech. J. \textbf{27}, 379--423 (1948);
\textbf{27}, 623--656 (1948).

\bibitem{PurvisLahav2013Cell}
J.~E.~Purvis and G.~Lahav,
\emph{Encoding and decoding cellular information through signaling dynamics},
Cell \textbf{152}(5), 945--956 (2013).
doi:10.1016/j.cell.2013.02.005

\bibitem{Marshall1995Cell}
C.~J.~Marshall,
\emph{Specificity of receptor tyrosine kinase signaling: transient versus sustained extracellular signal-regulated kinase activation},
Cell \textbf{80}(2), 179--185 (1995).
doi:10.1016/0092-8674(95)90401-8

\bibitem{Traverse1992BiochemJ}
S.~Traverse, N.~Gomez, H.~Paterson, C.~Marshall, and P.~Cohen,
\emph{Sustained activation of the mitogen-activated protein (MAP) kinase cascade may be required for differentiation of PC12 cells. Comparison of the effects of nerve growth factor and epidermal growth factor},
Biochem. J. \textbf{288}(2), 351--355 (1992).
doi:10.1042/bj2880351

\bibitem{Cowley1994Cell}
S.~Cowley, H.~Paterson, P.~Kemp, and C.~J.~Marshall,
\emph{Activation of MAP kinase kinase is necessary and sufficient for PC12 differentiation and for transformation of NIH 3T3 cells},
Cell \textbf{77}(6), 841--852 (1994).
doi:10.1016/0092-8674(94)90133-3

\bibitem{ItoSagawa2013}
S.~Ito and T.~Sagawa,
\emph{Information thermodynamics on causal networks},
Phys. Rev. Lett. \textbf{111}, 180603 (2013).
doi:10.1103/PhysRevLett.111.180603

\bibitem{ZumsandeGross2010}
M.~Zumsande and T.~Gross,
\emph{Bifurcations and chaos in the MAPK signaling cascade},
J. Theor. Biol. \textbf{265}(3), 481--491 (2010).
doi:10.1016/j.jtbi.2010.04.025

\bibitem{Seifert2005}
U.~Seifert,
\emph{Entropy production along a stochastic trajectory and an integral fluctuation theorem},
Phys. Rev. Lett. \textbf{95}, 040602 (2005).
doi:10.1103/PhysRevLett.95.040602

\bibitem{Crooks1999}
G.~E.~Crooks,
\emph{Entropy production fluctuation theorem and the nonequilibrium work relation for free energy differences},
Phys. Rev. E \textbf{60}(3), 2721--2726 (1999).
doi:10.1103/PhysRevE.60.2721

\bibitem{Gaspard2004}
P.~Gaspard,
\emph{Fluctuation theorem for nonequilibrium reactions},
J. Chem. Phys. \textbf{120}(19), 8898--8905 (2004).
doi:10.1063/1.1688758

\bibitem{Collin2005}
D.~Collin \emph{et al.},
\emph{Verification of the Crooks fluctuation theorem and recovery of RNA folding free energies},
Nature \textbf{437}(7056), 231--234 (2005).
doi:10.1038/nature04061

\bibitem{Sagawa2014}
T.~Sagawa \emph{et al.},
\emph{Single-cell \emph{E.\ coli} response to an instantaneously applied chemotactic signal},
Biophys. J. \textbf{107}(3), 730--739 (2014).
doi:10.1016/j.bpj.2014.06.017

\bibitem{LebowitzSpohn1999}
J.~L.~Lebowitz and H.~Spohn,
\emph{A Gallavotti--Cohen-type symmetry in the large deviation functional for stochastic dynamics},
J. Stat. Phys. \textbf{95}, 333--365 (1999).

\bibitem{SeifertReview2012}
U.~Seifert,
\emph{Stochastic thermodynamics, fluctuation theorems and molecular machines},
Rep. Prog. Phys. \textbf{75}, 126001 (2012).
doi:10.1088/0034-4885/75/12/126001

\bibitem{Zhou2011PLOS}
J.-P.~Zhou \emph{et al.},
\emph{Systems Biology Modeling Reveals a Possible Mechanism of the Tumor Cell Death upon Oncogene Inactivation in EGFR Addicted Cancers},
PLOS ONE \textbf{6}(12), e28930 (2011).
doi:10.1371/journal.pone.0028930

\bibitem{Finch2012CellSig}
A.~R.~Finch, C.~J.~Caunt, R.~M.~Perrett, K.~Tsaneva-Atanasova, and C.~A.~McArdle,
\emph{Dual specificity phosphatases 10 and 16 are positive regulators of EGF-stimulated ERK activity: Indirect regulation of ERK signals by JNK/p38 selective MAPK phosphatases},
Cellular Signalling \textbf{24}(5), 1002--1011 (2012).
doi:10.1016/j.cellsig.2011.12.021

\end{thebibliography}
\end{document}